# FUZZY CONTROLLER FOR MATRIX CONVERTER SYSTEM TO IMPROVE ITS QUALITY OF OUTPUT

[1]N.Mahendran*, [2]Dr.G.Gurusamy

[1]Research Scholar, Department of Electrical Engineering, Bannari Amman Institute of Technology, Sathyamangalam, Erode District, India.*E-mail:
mahendrann76@rediffmail.com

[2] Dean, Department of Electrical Engineering, Bannari Amman Institute of Technology, Sathyamangalam, Erode District, India.[2]E-mail:
magi_n2@rediffmail.com

*Corresponding Author

*Abstract:* *In this paper, Fuzzy Logic controller is developed for ac/ac Matrix Converter. Furthermore, Total Harmonic Distortion is reduced significantly. Space Vector Algorithm is a method to improve power quality of the converter output. But its quality is limited to 86.7%.We are introduced a Cross coupled DQ axis controller to improve power quality. The Matrix Converter is an attractive topology for High voltage transformation ratio. A Matlab / Simulink simulation analysis of the Matrix Converter system is provided. The design and implementation of fuzzy controlled Matrix Converter is described. This AC-AC system is proposed as an effective replacement for the conventional AC-DC-AC system which employs a two-step power conversion.*

## 1. INTRODUCTION

Real development of matrix converters starts with the work of Venturini and Alesina published in 1980. They presented the power circuit of the converter as a matrix of bi-directional power switches and they introduced the name "Matrix Converter." One of their main contributions is the development of a rigorous mathematical analysis to describe the low-frequency behavior of the converter, introducing the "low frequency modulation matrix" concept. In their modulation method, also known as the direct transfer function approach, the output voltages are obtained by the multiplication of the modulation (also called transfer) matrix with the input voltages. A conceptually different control technique based on the "fictitious dc link" idea was introduced by Rodriguez in 1983.
In this method the switching is arranged so that each output line is switched between the most positive and most negative input lines using a PWM technique, as conventionally used in standard voltage source inverters. This concept is also known as the "indirect transfer function" approach. In 1985/86, Ziogas et al published 2 papers which expanded on the fictitious dc link idea of Rodriguez and provided a rigorous mathematical explanation. In 1983 Braun and in 1985 Kastner and Rodrigue introduced the use of space vectors in the analysis and control of matrix converters. In 1989, was published the first of a series of the papers in which principles of the Space Vector Modulation (SVM) were applied to the matrix converter with the modulation problem. The modulation methods based on the Venturini approach, are known as "direct methods", while those based on the fictitious dc link are known as "indirect methods" It was experimentally confirmed by Kastner and Rodriguez in 1985 and Neft and Schauder in 1992 that a matrix converter with only 9 switches can be effectively used in the vector control of an induction motor with high quality input and output currents. However, the simultaneous commutation of controlled bidirectional switches used in Matrix Converters is very difficult to achieve without generating over current or overvoltage spikes that can destroy the power semiconductors. This fact limited the practical implementation and negatively affected the interest in Matrix Converters. Fortunately, this major problem has been solved with the development of several multistep commutation strategies that allow safe operation of the switches. In 1989 Burany introduced the later named "semi-soft current commutation" technique. Other interesting commutation strategies were introduced by Ziegler et al and Clare and Wheeler in 1998.

Total Harmonic Distortion (THD) is the most common power quality index to describe the quality of power electronic converter. In general, all the output voltage of power electronic converters is not purely sinusoidal. The THD of the output voltage can be defined as:

$$\text{THD} = \frac{\sqrt{V_2^2 + V_3^2 + V_4^2 + \cdots\cdots + V_n^2}}{V_1} \qquad (1)$$

Where n denotes the harmonic order and 1 is the fundamental quantity. For inverter application, THD represents how close the ac output waveform with pure sinusoidal waveform. A High quality matrix converter system should have low THD. Various study has been made on harmonic losses at electrical machine, it reveals that, presence of harmonic current in winding causes an increased copper loss [1]. The stator copper loss on a non sinusoidal supply is proportional to the square of the total rms current [2]. The core loss in the machine is increased by the presence of harmonics in the supply voltage and current. Magnitude of harmonic loss obviously depends upon the harmonic content of the motor voltage and current. Large harmonic voltage at low harmonic frequencies cause significantly increased machine loss and reduced efficiency [3]. The third harmonic injection scheme for the three phase diode rectifier for reducing the harmonic currents has drawn some promising results. A space vector based PWM strategy that closely approximates in the switching angles of the selective harmonic elimination PWM strategy [5]. While drawing sinusoidal input currents with unity power factor from the ac source, and having high power density and efficiency. In order to get pure sinusoidal wave either we have to design a filter or converter with different control techniques. Filter design for high rated machine includes weight and size of the total systems [6]. It affects the power processing capabilities. Matrix converter is a single stage converter and they need no energy storage components except small input ac filters for elimination of switching ripples [4]. However, a practical industrial application is still limited and the modulation method for the matrix converter is also understood to limited engineering people because of the high level of complexity and limited materials to explain its operating principle easily [7]. Meanwhile, the standard voltage source inverter (VSI) and its relevant space vector modulation (SVM) are well known to many engineering people due to the opposite reasons. Therefore, it would be a good approach to explain the operating principle of the matrix converter by adopting standard SVM concept. The first modulator was proposed by Venturini and he used a complicated scalar model that gave a maximum voltage transfer ratio of 0.5. An injection of a third harmonic of the input and output voltage was proposed in order to fit the reference output voltage in the input voltage system envelope, and the voltage transfer ratio reached the maximum value of 0.86. The objective of this research is to propose a PWM strategy to reduce THD, which is reported to R, RL and Motor load. The THD is investigated in the matrix converter fed ASD [9], [10]. Matrix converter controller is modified with the addition of a fuzzy controller and thereby a new fuzzy controller is proposed together with Space Vector PWM strategy. These results are compared with various operating frequencies.

## 2. MATRIX CONVERTER OPERATION

The ac/ac converters are commonly classified into indirect converter which utilizes a dc link between the two ac systems and direct converter that provides direct conversion. Indirect converter consists of two converter stages and energy storage element, which convert input ac to dc and then reconverting dc back to output ac with variable amplitude and frequency. The operation of these converter stages is decoupled on an instantaneous basis by means of energy storage element and controlled independently, so long as the average energy flow is equal. Therefore, the instantaneous power flow does not have to equal the instantaneous power output. The difference between the instantaneous input and output power must be absorbed or delivered by an energy storage element within the converter. The energy storage element can be either a capacitor or an inductor. However, the energy storage element is not needed in direct converter. In General, direct converter can be identified as three distinct topological approaches. The first and simplest topology can be used to change the amplitude of an ac waveform. It is known as an ac controller and functions by simply chopping symmetric notches out of the input waveform. The second can be utilized if the output frequency is much lower than the input source frequency. This topology is called a Cycloconverter, and it approximates the desired output waveform by synthesizing it from pieces of the input waveform. The last is matrix converter and it is most versatile without any limits on the output frequency and amplitude. It replaces the multiple conversion stages and the intermediate energy storage element by a single power conversion stage, and uses a matrix of semiconductor bidirectional switches, with a switch connected between each input terminal to each output terminal as shown in Fig 1. With this general arrangement of switches, the power flow through the converter can reverse. Because of the absence of any energy storage element, the instantaneous power input must be equal to the power output, assuming idealized zero-loss switches. However, the reactive power input does not have to equal the reactive power output. It can be said again that the phase angle between the voltages and currents at the input can be controlled and does not have to be the same as at the output.

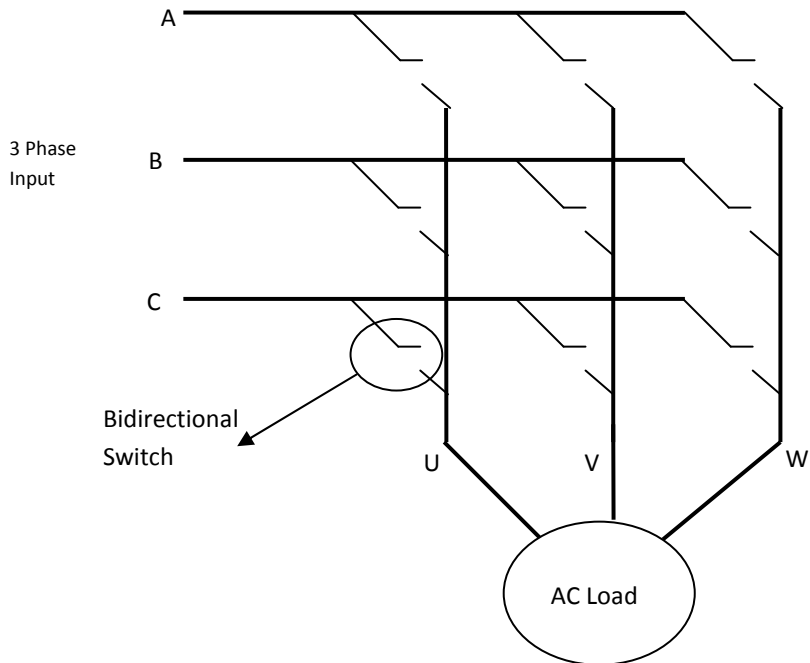

Fig. 1 Matrix converter with 3 phase AC load

## 3. CONTROL PRINCIPLE

A sinusoidal ac voltage source having an amplitude $V_1$ and angular frequency $w_1$ ($2\Pi f_1$) is connected to the input terminal of matrix converter. This is applied sinusoidal voltage is converted into an output voltage with amplitude of $V_0$ and angular frequency $w_0$ ($2\Pi f_0$), which is applied to the load. Upper limit of the range of variation of the output frequency lies at a point lower than the input frequency.

$$V_{ABC} = K_i e^{j(w_i t + \Phi_i)} \quad (2)$$
$$V_{UVW} = K_0 e^{j(w_0 t + \Phi_0)} \quad (3)$$

Control principle(S (t))

$$S(t) = \frac{K_0 e^{j(w_0 t + \phi_0)}}{K_i e^{j(w_i t + \phi_i)}} \quad (4)$$

Where, $\Phi_i$ = Input phase displacement, $\Phi_0$ = Output phase displacement

Maximum voltage transfer ratio 50% is possible in Venturini Algorithm. An improvement in the achievable voltage ratio to $\frac{\sqrt{3}}{2}$ (for 87%) is possible by adding common mode voltage to the target output. We can investigate these voltage transformation ratio based on following theoretical analysis. There are different modulation techniques available from Venturini invention (1980).Implementation of venturini algorithm is difficult calculation. We are looking forward towards simple algorithm and improved voltage transformation ratio. Following switching equations will explain about output voltage relationship and various switching techniques.

**3.1 Venturini modulation method (Venturini first method)**

$$m_{kj} = \frac{t_{kj}}{T_{sep}} = \frac{1}{3}[1 + \frac{2v_k v_j}{v_{im}^2}], \text{ for k=A, B, C and j=a, b, c} \quad (5)$$

Where, $m_{kj}$ = Modulation duty cycle

Voltage transformation ratio is 50% [12].From equation (5) second term q=1/2.

### 3.2 Venturini Optimum method (Venturini Second method)
Employs common mode addition and maximum transformation ratio is 87% [11]. This is also known as displacement factor control. Displacement factor control can be introduced by inserting a phase shift between the measure input voltages and inserted voltage ($v_k$).

$$m_{kj} = \frac{t_{kj}}{T_{sep}} = \frac{1}{3}[1 + \frac{2v_k v_j}{v_{im}^2} + \frac{4q}{3\sqrt{3}}\sin(\omega_i t + \beta_k)\sin(3\omega_i t)], \text{ For k=A, B, C and j=a, b, c}$$

$\beta_k = 0, 2\Pi/3, 4\Pi/3$ for k=A, B, C  (6)

Where, $V_{im}$= Maximum Input voltage

### 3.3 Scalar Modulation method:
Actuation signals are calculated directly from measurement of input voltages. Voltage transformation ratio is 87%.

$$m_{kj} = \frac{t_{kj}}{T_{sep}} = \frac{1}{3}[1 + \frac{2v_k v_j}{v_{im}^2} + \frac{2}{3}\sin(\omega_i t + \beta_k)\sin(3\omega_i t)] \quad (7)$$

This method yields virtually identical switch timings to the optimum Venturini method. The maximum output voltage (q=$\frac{\sqrt{3}}{2}$) are identical. Only the difference between the methods is that the right most term addition is taken pro rata with q in the Venturini method.

### 3.4 SPVM
Space vector pulse width modulation is applied to output voltage and input current control. This method is advantage because of increased flexibility in choice of switching vector for both input current and output voltage control can yield useful advantage under unbalanced conditions [9].The three phase variables are expressed in space vectors. For a sufficiently small time interval, the reference voltage vector can be approximated by a set of stationary vectors generated by a matrix converter. If this time interval is the sample time for converter control then, at the next sample instant when the reference voltage vector rotates to a new angular position, it may correspond to a new set of stationary voltage vectors. Carrying this process onwards by sampling the entire waveform of the desired voltage vector being synthesized in sequence, the average output voltage would closely emulate the reference voltage. Meanwhile, the selected stationary vectors can also give the desirable phase shift between input voltage and current. The modulation process thus required consists of two main parts: selection of the switching vectors and computation of the vector time intervals. From the above methods give the theoretical maximum voltage gain of 0.866 though they use different approaches. This is realized in Venturini method, by adding third harmonic components of both input and output voltages to the desired output waveform whereas in the SVM method. Modulation of line to line voltage naturally gives an extended output voltage capability. The computational procedure required by SVM method is less complex than that for Venturini method because of the reduced number of sine function computations. The number of switch commutations per switching cycle for SVM method is 20% less than that of Venturini method, having 7 switch commutations as opposed to 9 [8].

### 3.5 Indirect Modulation method
This method aims to increase the maximum voltage ratio above 86.6%limit of other methods.

$$v_o = (Av_i)B = \frac{3K_A K_B V_{im}}{2}\begin{bmatrix}\cos(\omega_i t)\\ \cos(\omega_i t + 2\pi/3)\\ \cos(\omega_i t + 45\pi/3)\end{bmatrix} \quad (8)$$

Voltage ratio q=$3K_A K_B$ /2.Clearly A and B modulation steps are not continuous in time as shown above.

$K_A$ = 2sqrt (3) $V_{im}$/$\Pi$  (9)

$K_B = 2/\Pi$ (10)

Then, $q = 6\sqrt{3}/\Pi^2 = 105.3\%$ (11)

The voltage output is greater than the previous method. For the values q>0.866, the mean output voltage no longer equals the target output voltage in each switching interval. This inevitably leads to low frequency distortion in the output voltage and /or the input current compared to other methods with q<0.866. For q<0.866, the indirect method yields very similar results to the direct methods.

## 4.0 CROSS COUPLED FUZZY CONTROLLER

Most widely used method of testing a fuzzy controller design is by simulation. In this research work Fuzzy Logic Toolbox in MATLAB Simulink is used to solve the problem shown in Figure 2.

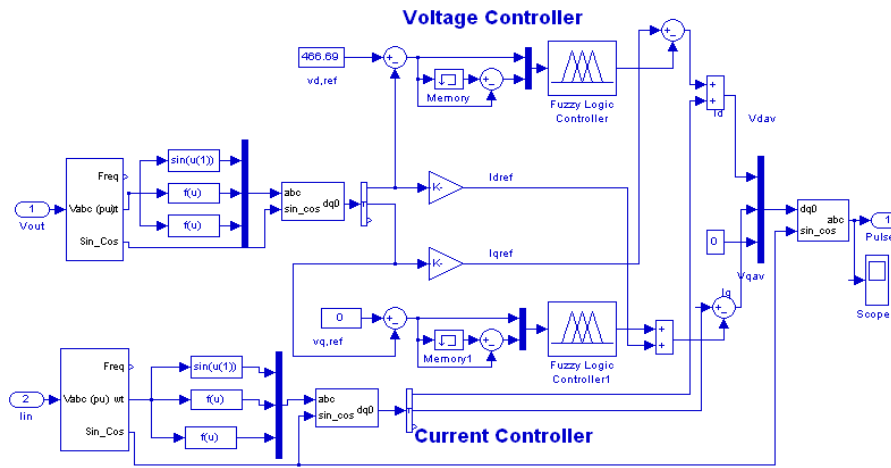

Fig. 2 Cross coupled fuzzy structure for matrix converter

This toolbox can simulate in various states such as transient state response and steady state error, corresponding to each control goal. Different combination of input can easily be tested to observe the corresponding output. Fig 2 shows the flow of the typical design procedure used to develop a fuzzy structure. The simulation and testing are conducted several times until a satisfactory result is accomplished. The results are refined by parameter tunings that are based on intuitive experiences and the qualitative results obtained from time to time. Each universe of discourse for two inputs and an output is divided into seven fuzzy subsets that consists of negative logic(nl),negative medium(nm),negative small(ns), zero(z),positive small(ps),positive medium(pm)and positive large(pl). The membership function chosen are the classical triangular shape of 50% overlap. The portion of fuzzy subsets and shapes of Membership functions are shown in fig 3 and fig 4. Rule base is derived based on the characteristics of the RMS value of the output signal that is similar to the response for a second order system by applying step input. If rms value of the load voltage is less than the rms value of the reference voltage at point, then error signal is positive consequently, the control action has to be increased, thus giving positive 'ce' to enable the load voltage to reach the set point. The combinations of inputs and control action are summarized in table1.The inference method of Mamdani is max-min composition is chosen in the work to simplify the programming algorithm. After several trials has been made to select membership function. And finally it is decided to select triangular membership function.

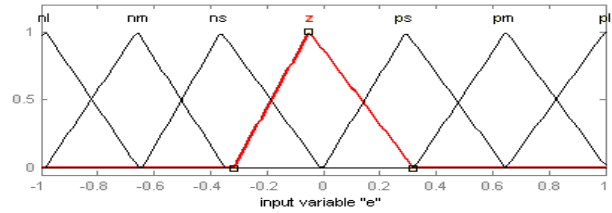

Fig. 3.Input membership function

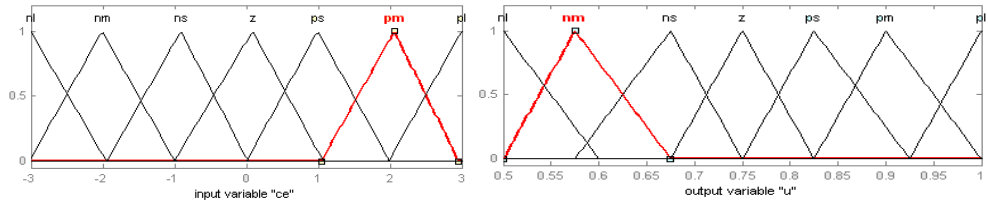

Fig. 4.Input ce and Output variable u membership function

Table 1 Rule base for Fuzzy controller

|   |   | Change in error(ce) | | | | | | |
|---|---|---|---|---|---|---|---|---|
|   |   | NB | NM | NS | ZE | PS | PM | PB |
| E | NB | NB | NB | NM | NM | NS | NS | ZE |
| R | NM | NB | NM | NM | NS | NS | ZE | PS |
| R | NS | NM | NM | NS | NS | ZE | PS | PS |
| O | ZE | NM | NS | NS | ZE | PS | PS | PM |
| R | PS | NS | NS | ZE | PS | PS | PM | PM |
| (e) | PM | NS | ZE | PS | PS | PM | PM | PB |
|   | PB | ZE | PS | PS | PM | PM | PB | PB |

The voltage command is then compared to a triangular carrier at the desirable switching frequency. In this work voltage command is tracked by fuzzy controller and compared with current controller as shown in fig 2.

Computations of d and q axis parameters are shown in equation 12 to equation 14.

Computation of dq axis

$$V_d = \frac{2}{3}\left[V_A \sin wt + V_B \sin(wt - \frac{2\pi}{3}) + V_C \sin(wt + \frac{2\pi}{3})\right] \quad (12)$$

$$V_q = \frac{2}{3}\left[V_A \cos wt + V_B \cos(wt - \frac{2\pi}{3}) + V_C \cos(wt + \frac{2\pi}{3})\right] \quad (13)$$

$$V_0 = \frac{1}{3}\left[V_A + V_B + V_C\right] \quad (14)$$

Where w=rotation speed in rad/s

Computation of dq to abc

$$V_A = \left[V_d \sin wt + V_q \cos(wt) + V_0)\right] \quad (15)$$

$$V_B = \left[V_d \sin(wt - \frac{2\pi}{3}) + V_q \cos(wt - \frac{2\pi}{3}) + V_0\right] \quad (16)$$

$$V_C = \left[V_d \sin(wt + \frac{2\pi}{3}) + V_q \cos(wt + \frac{2\pi}{3}) + V_0\right] \quad (17)$$

Where w=rotation speed in rad/s
Input contains vectorized signal of $V_d$, Vq and $V_o$ components.

## 4. RESULTS AND ANALYSIS

In order to analyze the performance of the proposed cross coupled dq axis controller using fuzzy logic controller, a simulation mode of the Matrix converter was implemented in Sim power systems from Simulink using ideal switches.

A three phase star connected RL load with R=10Ω and L=200µH were used in the simulation. The switching frequency of Matrix converter is 16 KHz and input voltage is 440V, 60 Hz. The Modulation index is dynamically corrected by the fuzzy controller and the output current is almost sinusoidal and balanced.

From the fig 5 to fig 8 shows the difference between input parameters and output parameters for 30Hz fixed frequency. Compare input voltage with output voltage to read about voltage transformation ratio. This can be possible by this proposed method. Here output parameters are measured without output filters. This parameter can be still shaped by using output filters. PQ settings are linearly varies each other as shown in fig 9 and fig 10. For same RL load different values of frequency settings are made and results were obtained for 30Hz and 60Hz. We proved that this measured value is high quality output.

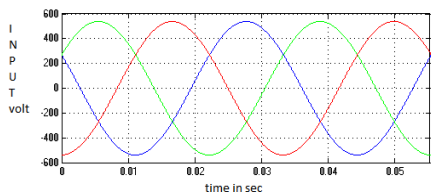
Fig. 5 Input Voltage (30Hz)

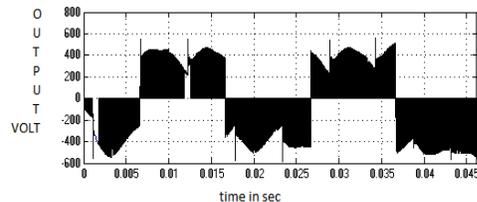
Fig. 6 output voltage (30Hz)

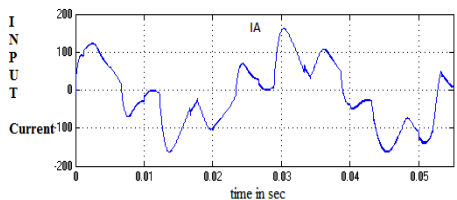
Fig. 7 Input Current (30Hz, per phase)

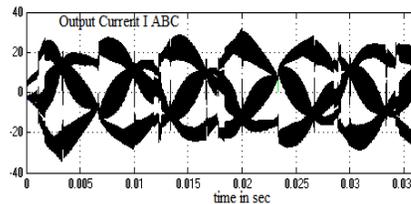
Fig. 8 output current (30Hz)

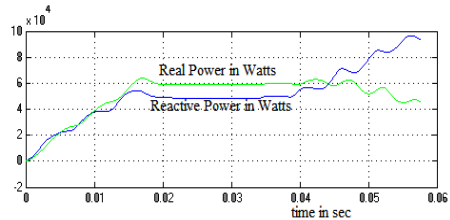
Fig. 9 Input PQ setting (30Hz)

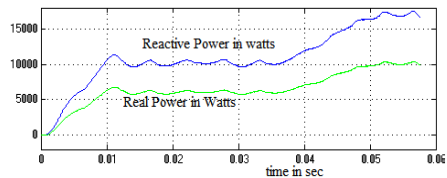
Fig.10. Output PQ setting (30Hz)

Change in frequency normally makes lot of difference in output side. Here we have done simulation work for 30Hz and 60Hz. PQ setting for power system load was measured as shown in fig 15 and fig 16. From Fig 11 to Fig 14 shows variation of input and output voltage for 60Hz.

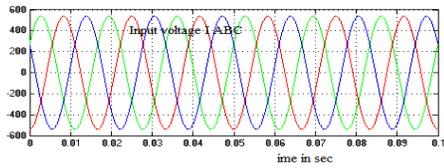
Fig. 11 Input voltage (60Hz)

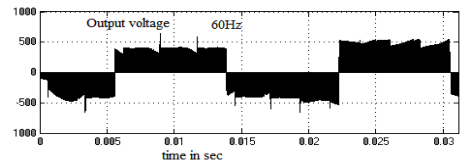
Fig. 12 Output Voltage (60Hz)

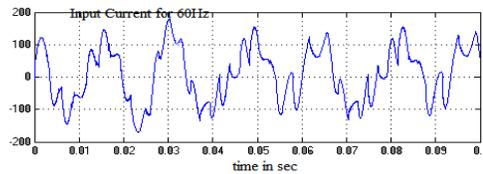
Fig. 13 Input current (60Hz)

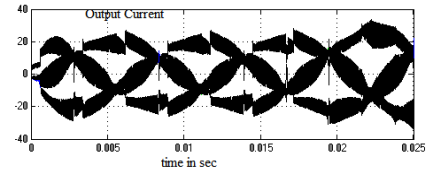
Fig. 14 Output Current (60Hz)

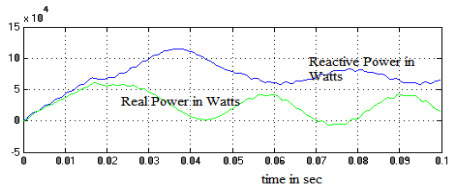
Fig. 15. Input PQ Setting (60Hz)

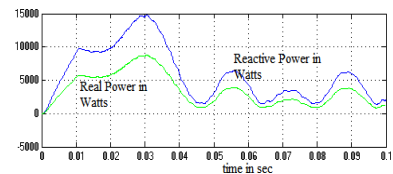
Fig. 16. Output PQ Setting (60Hz)

Table 2 shows the open loop simulation of matrix converter with different load conditions. THD are measured for open loop Space vector control method. Space Vector Modulation provides excellent output performance, optimized efficiency, and high reliability compared to similar inverters with conventional Pulse Width Modulation.

Table 2 THD comparision R,RL and Motor load(Open loop simulation),without output filter(Parameters as shown in Appendix A )

| Parameter | R Load | RL Load | Motor Load |
|---|---|---|---|
| Switching Frequency | 16KHz | 16KHz | 16KHz |
| Output THD | 65.64% | 22.05% | 4.9% |
| Input THD | 22.7% | 22.59% | 22.43% |

From the above analysis it seems that voltage transformation ratio of the matrix convertere is almost reaches to maximum level . Table 2 shows measured THD for different load conditions.

## 5. CONCLUSION

In this paper a new and simple cross coupled dq axis controller for MC was proposed. Voltage control was implemented using fuzzy logic controller and space vector modulation .Two similar fuzzy controller is implemented to get cross coupled structure. This make synchronization of D and Q axis parameter not much complicated. Implementation in hardware is easy.  The fuzzy controller forces the amplitude of the output current space vector to be constant so that the output current is free of harmonic. The same algorithm can be implanted for different converter.

## REFERENCES

[1]     T.V.Avadhanly, Derating factors of 3phase IM, *J.Inst Eng (India)*, April1974, pp.113-117.

[2]     V.B.Honsinger, Induction motors operating from inverters, *Conference of Rec.IEEE Industrial Applications Society Annual meeting*, 1980, pp.1276-1285.

[3]     JMD.Murphy, VB.Honsinger, Efficiency optimization of inverter fed induction motor drives, *Conference of Rec.IEEE Industrial Applications Society Annual meeting*, 1982, pp.544-552.

[4]     AC Williamson, The effects of system harmonics upon machines, *Int J.Electrical Engineering Education,* April 1982, pp.145-155.

[5]     P.D.Ziogas,S.I.Khan and M.H. Rashid ,Analysis and design of forced commutated cycloconmverter structures with improved transfer characteristics, *IEEE transactions on Industrial Electronics*,vol.IE-33, August 1986,pp.271-280.

[6]     L. Huber,D.Borojevie and N.Burany , Voltage space vector based PWM control of forced commutated cycloconvertors,  *proc.IEEE IECON'89*, 1989,pp.106-111.

[7]     Patrick W.Wheeler,Jose Rodriguez,Jon C.Clare ,Lee Empreingham and Alejandro Weinstein ,Matrix Converters: A Technology Review, *IEEE Transactions on Industrial Electronics*,Vol.49,No.2,April 2002,pp.282-285

[8]     G.K.Sing ,A research survey of induction motors operation with non sinusoidal supply waveforms, *Electric power systems research*, Elsevier, 75, 2005, pp. 200-213.

[9]     C.L.Weft and C.D.Schauder,Theory and design of 30HP matrix converter, *IEEE Transactions on Industrial Applications*, Vol.28, No 3, 2006, pp.546-551.

[10]    N.Mahendran and Dr.G.Gurusamy ,THD analysis of matrix converter fed load, *IEEE Eighth International Conference on Power Electronics and Drive Systems (PEDS 2009）*, Taiwan, 2009, pp.829-832.

[11]    C Watthanasarn, L. Zhang, D T W Liang, Anslysis and DSP based Implementation of Modulation Algorithms for AC to AC matrix converter, *IEEE 1996*, pp.1053-1058

[12]    M B B Venturini and A. Alesina, A New Sine  Wave In, Sine Wave Out conversion technique Eliminates Reactive Elements, *Proceedings of Powercon7*, 1980.

**Appendix A**
Design Specifications of the proposed approach (Power system load)

| Parameter | Rating |
|---|---|
| Load Resistance | 10Ω |
| Line Inductance | 200µH |
| Device Switching Frequency | 16KHz |

Design Specifications of Motor Load (Open loop configuration)

$V_{ll}$=220V, F=60Hz, $R_s$=0.435Ω, $L_s$=2mH, $R_r$=0.816 Ω, $L_r$=2mH, $L_m$=69.31mH

Where,

$V_{ll}$=Line to Line voltage, $R_s$=Stator Resistance, $L_s$= Stator Inductance, $R_r$=Rotor Resistance and $L_r$=Rotor Inductance

**Appendix B**
Input filter Design

Input filter Capacitor: $C_f$

$$C_f = \frac{2 \times P}{3V_m^2 \omega_i}$$, Where P=Power Rating, Vm=Peak of Input voltage and $\omega_i$=angular input Frequency.

$$L_f = \frac{1}{(2\pi f_c)^2 c_f}$$, Where $f_c$=cut off frequency lower than the switching frequency

## BIOGRAPHICAL NOTES:

*N.Mahendran* received his BE in Electrical and Electronics Engineering from the Madras University in 2000, his M.Tech in Control Systems and Instrumentation Engineering from the University of SASTRA, Thanjavur in 2004 and he is pursuing PhD in Electrical Engineering from the Anna University-Chennai. He is currently a Research Scholar of Electrical Engineering at Bannari Amman Institute of Technology and his primary profession is teaching at Maha college of Engineering, Salem. His research interests include machine controls, electric drives, and Artificial intelligence and control systems.

*Dr.G.Gurusamy* obtained his Pre University education at St John's College, Palayankottai and Trinelvelli district. He joined PSG College of Technology, Coimbatore, in the year 1962 to pursue his Engineering course. He was graduated in Electrical Engineering in 1967. Latter he obtained his M.E. (Applied Electronics) in 1972 and Ph.D in Control Systems in 1983. He has more than 30 years of teaching Experience in PSG college of Technology. He is currently working as Dean of Department of Electrical and Electronics Engineering in Bannari Amman Institute of Technology, Sathyamangalam. His fields of interest are advanced control, Power Quality, Digital control, Optimization and Bio medical Electronics.